\documentclass{article}
\usepackage{graphicx} 

 \usepackage[numbers]{natbib}
\usepackage {amsmath}
\usepackage{xcolor}
\usepackage{tabulary}
\usepackage{colortbl}

\title{Salience-based stakeholder selection to maintain stakeholder coverage in solving the next release problem
}                      
\author{IM del \'Aguila   \& J. del Sagrado \\
\small{University of Almer\'ia. Spain}  \\
\small{Accepted Information \& software technology }\\
}

\date{}

\begin{document}

\maketitle

\begin{abstract}
\noindent\textbf{Context:}  The quantification of stakeholders plays a fundamental role in the selection of appropriate requirements, as their judgement is a significant criterion, as not all stakeholders are equally important. The original proposals modelled stakeholder importance using a weighting approach that may not capture all the dimensions of stakeholder importance. 
Furthermore, actual projects involve a multitude of stakeholders, making it difficult to consider and compute all their weights. 
 These facts lead us to search for strategies to adequately assess the importance concept, reducing the elicitation effort.

\noindent\textbf{Objective:} We propose grouping strategies as a means of reducing the number of stakeholders to manage in requirement selection while maintaining adequate stakeholder coverage (how selection meets stakeholder demands).

\noindent\textbf{Methods:} Our approach is based on the salience of stakeholders, defined in terms of their power, legitimacy, and urgency. Diverse strategies are applied to select important stakeholder groups. We use \emph{k-means}, \emph{k-medoids}, and \emph{hierarchical} clustering, after deciding the number of clusters based on validation indices.

\noindent\textbf{Results:} 
Each technique found a different group of important stakeholders. The number of stakeholder groups suggested experimentally (3 or 4) coincides with those indicated by the literature as definitive, dominant, dependent, and dangerous for 4 groups; or critical, major, and minor for 3 groups. 
Either for all the stakeholders and for each important group, several requirements selection optimisation problems are solved.  The tests do not find significant differences in coverage when important stakeholders are filtered using clustering, regardless of the technique and number of groups, with a reduction
between 66.32\% and 87.75\% in the number of stakeholders considered. 

\noindent\textbf{Conclusions:} 
Applying clustering methods to data obtained from a project is useful in identifying the group of important stakeholders. 
The number of suggested groups matches the stakeholders' theory, and the stakeholder coverage values are kept in the requirement selection.

\end{abstract}

\maketitle

\section{Introduction}

To build a useful technological system, we need to know its requirements; to know its requirements, we need to know the desires and needs of stakeholders~\cite{glinz2007guest}. 
That is, eliciting, documenting, and validating stakeholders' requirements are fundamental activities. However, due to the variety of stakeholders' backgrounds and/or interests, their influence on Requirements Engineering planning decisions can be considered a problem in itself. In fact, stakeholder theory is a well-established research area~\cite{harrison2019cambridge},  where the problem of identifying the right stakeholders is a basic task that precedes any other system development activity~\cite{mitchell2019stakeholder, pacheco2012}. This task is usually based on "the degree to which managers give priority to competing stakeholders' claims in their decision-making process", called stakeholders' salience~\cite{mitchell2019stakeholder},  to define the importance of each stakeholder.

A stakeholder in an organisation is any group or individual who can affect or is affected by the achievement of the organisation's objective. Stakeholder theory posits that understanding and meeting the needs of various stakeholders is essential for the survival of businesses~\cite{freeman1984}.
Specifically, the development of software systems to support organisation processes also involves the management of stakeholders' needs, both business and technical, to produce a software application that fulfils their needs.   
Stakeholder needs have to be translated into a formal set of stakeholders' requirements. 
These requirements represent conditions or capabilities that must be included in the next release of the current software project~\cite{babok2015,SWEBOK2014}.

However, not all stakeholders' requests can be included in the next software to be delivered. On the one hand, due to the limited amount of available resources, and on the other hand, because each stakeholder has a different significance in the organisation and a different importance (that is, the priority assigned by the managers, also called salience) for the project. 
If inappropriate stakeholders were identified or unimportant requirements were selected, it could lead to a not-successful software system because it will not cover the real needs. That is, once stakeholders have been identified, their quantification and importance arrangement play a basic role in selecting the appropriated requirements because their judgment is one of the main criteria in the requirement selection.

In fact, from its original proposal~\cite{bagnall2001}, the next release problem (\ensuremath{\mathsf{NRP}}) is a complex multi-criteria decision process that entails achieving a balance between the value that requirements add to the project outcome (called satisfaction) and their cost~\cite{zhang2007, sagrado2015, dominguez2019, zhang2018}. It starts from the values that each stakeholder estimates as their benefit from the inclusion of the given requirements. However, stakeholders are not equally important, so these values are affected by a weight that measures the significance of each stakeholder, reinforcing the need to establish a stakeholder quantification step before doing the requirement selection process. As more stakeholders in a project need to be managed, the more complex \ensuremath{\mathsf{NRP}} becomes. 
Most cases use the weighted sum of values estimated by each stakeholder as a requirement outcome (i.e., satisfaction). Satisfaction is used as an optimisation objective to maximise, usually in combination with the development effort as cost that has to be minimised~\cite{bagnall2001,zhang2007}. 

This work has a twofold goal, first, to check if the available quantitative records of stakeholders in a software project are a valid source for defining stakeholders' weight in \ensuremath{\mathsf{NRP}} that matches the salience concept, which categorises  stakeholders based on the estimates of their three attributes: power, legitimacy, and urgency~\cite{mitchell1997}. In addition, we propose various approaches to group stakeholders based on clustering methods to check whether the selection of requirements, considering only the most important stakeholder group, is enough to uniformly cover all of them  (defining coverage as the percentage of stakeholder proposals / desires included in the set of selected requirements \cite{aguila2016}). That is, the management of a few stakeholders in \ensuremath{\mathsf{NRP}} gives adequate coverage to the total amount of the community of stakeholders in the project.

Our improved next release problem-solving process starts by identifying requirements and stakeholders. Then, a task has to be performed to determine the definitive stakeholders.  This task has as objective the triage of the definitive stakeholders, which will be selected according to their salience, becoming the unique stakeholders to consider in \ensuremath{\mathsf{NRP}}.
At this point, the requirement selection problem can be set up and solved, using any previously validated algorithm, before the development of the new software version.

Research has been conducted stating this goal, which brings us to some research questions and, finally, carrying out some empirical studies using an open data set. These studies give us the information needed to answer the next research questions.  

\begin{itemize}

\item[\emph{RQ1}] Is it possible to use quantitatively defined stakeholder salience to identify the most important stakeholders (definitive stakeholders) for \ensuremath{\mathsf{NRP}}?

\emph{Rationale:} As the stakeholders involved in a requirement selection problem are not equally important, our aim is to measure their importance based on their salience value ~\cite{mitchell1997}, obtained from the available quantitative records of the stakeholders involved in a software project. However, most previous efforts to help identify "who and what counts" \cite{rahman2015,poplawska2015} employ approximate quantification. Additionally, this measure should serve as a basis for identifying the different stakeholder groups.

\item[\emph{RQ2}] Can clustering techniques define an appropriate number of stakeholder groups compatible with stakeholder salience theory?

\emph{Rationale:} The objective of RQ2 is the triage of stakeholders according to their salience, and the identification of definitive stakeholders. For this purpose, clustering methods are used and the recommended number of stakeholder groups is checked against that suggested by stakeholder categorisation theories~\cite{mitchell1997,babok2015,SWEBOK2014}. Using such methods allows the process of stakeholder identification to be automated. The stakeholders in the definitive stakeholder group are the only ones that will be considered in the requirements selection problem, decreasing the initial number of stakeholders to be managed.

\item[\emph{RQ3}] Is employing only important stakeholders sufficient to cover all stakeholders in a project in \ensuremath{\mathsf{NRP}}?

\emph{Rationale:} The objective of RQ3 is to test whether the reduction of stakeholders to only the definitive stakeholder group results in a simplification of the initial requirement selection problem while satisfying most of the demands of the other stakeholders. And in this way, the appropriate requirements are selected.

\end{itemize}

The remainder of the paper is structured as follows. Section 2 includes the approaches followed by other authors to deal with stakeholder quantification, including the contribution of our research.  Section 3 presents the necessary background knowledge on clustering. The basis of stakeholder salience is presented in Section 4. Section 5 describes the methodology followed, including the description of the data set used and how stakeholders can be arranged according to the salience distribution. In addition, section 5 also explains, through a roadmap, how the clustering algorithms have been handled.
Section 6 presents the analysis of the results and answers the research questions. 
The identified threats to validity are reported in Section 7. The last section presents the conclusions.

\section{Related works}
Small projects have fewer stakeholders, so you can manage \ensuremath{\mathsf{NRP}} easily. However, larger projects, such as Jenkins, Django, SonarQube, or Kodi (mainly open source projects), need to perform more complex requirement selections because of the huge stakeholders' community to be taken into account. 
This kind of project usually involves a large number of heterogeneous stakeholders with wide and not unified interests. They could be current clients who paid for the system, potential new clients, users, authorities, or developers who define new requirements or conditions to be managed in the project, sometimes totally different and conflicting.

Several works have been performed to determine who can affect or is affected by the system and how much they are concerned~\cite{pacheco2012}, but the authors of this review only reported that 27\% of the studies include an assessment of stakeholders. The stakeholder theory defines "who really counts" in a system based on the salience of stakeholders~\cite{freeman1984, mitchell1997}, defining different groups of stakeholders but without clearly managing a salience measure. 

Some works define a set of rules to identify and categorise the key stakeholder considering five levels of influence~\cite{rahman2015, prasanth2018}.
Other studies afford the stakeholder quantification process~\cite{HUJAINAH2019, mughal2018, babar2015, babar2014bi}, although, to our knowledge, none of them deals with the impact of different stakeholder quantification strategies on the requirement selection decision. For example, by addressing the biasedness problem while identifying and prioritising stakeholders~\cite{mughal2018}, but focussing on the validity of some requirements exercises performed by groups that communicate using or not using a social network without including release planning decisions.

Certain proposals apply a value-based approach to define a stakeholder quantification method. They use nine estimated factors / attributes (each calculated from 3 to 7 additional values) to give an aggregated quantitative value of the influence of stakeholders~\cite{babar2014bi,babar2015}, forcing managers to cope with many quantitative attributes. These aggregated values have also been used in clustering approaches to  stakeholders assessment; based on a bi-metrics fuzzy c-means method to obtain 2 or 3 stakeholder clusters~\cite{babar2014bi} or by applying semi-automatically tools to group stakeholders~\cite{HUJAINAH2019}. However, in addition to needing to treat and estimate numerous values, such as experience, management skill, domain knowledge, practise over the stakeholder's domain, or self-esteem or objectivity,  none of them studies the impact of their proposal in the requirements selection decision.   
Other solutions study the influence of individual stakeholders on requirement selection. However, they left aside the stakeholder salience theory, which characterises stakeholders based on power, legitimacy, and urgency.
Examples of this are proposals to identify and prioritise stakeholders in large-scale software projects with the aim of prioritising requirements~\cite{lim2011, lim2013}, where stakeholder weights are estimated using only social network recommendations, that is, stakeholder legitimacy; or those that use the complete stakeholder preference matrix for requirements to apply genetic clustering to find convergence between the preferences of stakeholders to define groups of stakeholders~\cite{reyad2021}.

The related results made more complex versions of \ensuremath{\mathsf{NRP}} including many data to manage the significance of stakeholders in the project~\cite{finkelstein2009, veerappa2011}, tangling the problem.
Clustering algorithms for requirements prioritisation have been applied using together requirements rates and stakeholder data, such as the number of hours online per day searching for groups of requirements to give them a median value for all requirements in a given cluster~\cite{veerappa2011}.  An investigation of fairness between stakeholders has been treated by including new optimisation objectives in \ensuremath{\mathsf{NRP}}. These objectives are based on the mean number of requirements fulfilled and the individual values assigned to the requirements by all stakeholders, and the optimisation objectives based on the standard deviation of these values~\cite{finkelstein2009}.

Our proposal is to group stakeholders based on salience, to reduce the number of stakeholders involved in a requirement selection process, while maintaining the agreed definition for \ensuremath{\mathsf{NRP}}, so as to facilitate the application of abundant and validated algorithms to solve this problem~\cite{geng2018, nuh2021}, but with fewer data to manage. 
 The metric we propose to quantify the value of salience is driven by stakeholder theory \cite{freeman1984,mitchell1997} (unlike other works~\cite{babar2014bi,babar2015,HUJAINAH2019}
which, although they quantify stakeholders, based on many stakeholder attributes, are outside this framework) and is not an estimated value based on levels of influence (as in the cases of the works\cite{rahman2015,prasanth2018}). Our method of stakeholder quantification based on salience allows us to identify the group of definitive stakeholders. This identification has a clear impact on the problem of requirement selection by reducing the workload; as with a smaller group of stakeholders, we can identify a set of requirements that satisfies the demands of the overall community of stakeholders. Other works, such as ~\cite{lim2011, lim2013, reyad2021,finkelstein2009, veerappa2011} focus
on the impact that stakeholder weight has on requirement selection, but do not accomplish any stakeholder classification task. Instead, they embedded all stakeholder data into the requirement selection problem.

\section{Background}

The goal of clustering methods is to discover groups of similar objects within a data set. Thus, clustering approaches require some methods for measuring the (dis)similarity (or distance) between objects, so that objects in the same group should be closer to each other and further than those in other groups. Common clustering approaches can be classified into one of the following categories:

\begin{itemize}
\item \emph{Partitioning methods} that divide the data into a pre-specified number of groups.  \emph{K-means} and \emph{K-medoids} (that is, the partition algorithm around the medoids) belong to this category \cite{macQueen1967}.

\emph{K-means} algorithm \cite{hartigan1979} is perhaps the most popular strategy for finding clusters in the data. Each cluster is represented by its \emph{centroid}, which is the mean of the observations assigned to the cluster. Initially, once the number $k$ of groups has been specified, $k$ observations are chosen as centroids for the clusters. Clusters are defined so that the distance between each observation in a cluster and its centroid is a minimum. 
Although \emph{K-means} is simple, fast and can handle large data sets, the number $k$ of clusters has to be specified in advance, and the results obtained depend on the initial centroids selection.  
Besides, it is sensitive to data ordering (i.e. if you rearrange your data, possibly you will get a different solution) and it is also sensitive to anomalous data points and outliers.

\item Partitioning Around Medoids (PAM) \cite{kaufman1990b} rests on the concept of \emph{medoid}, which is an observation within a cluster such that the sum of the distances between it and all other observations in the cluster is a minimum. 
Clusters are constructed by assigning each observation to the nearest medoid. Then, \emph{PAM} tries to improve the quality of the clustering by interchanging the medoids with the other observations and checking if the distances with respect to the medoid are reduced.  Although it is more reliable and less sensitive to outliers, \emph{PAM} requires more computational effort than \emph{K-means}.

\item \emph{Hierarchical methods} \cite{murtagh2012}. In contrast to partitioning methods (i.e., \emph{K-means} and PAM), \emph{hierarchical methods} group observations based on their similarity and does not require prespecifying the number of clusters to be produced. They return a dendogram, a tree-based representation of objects and groups.

The most representative method in this category is \emph{agglomerative clustering}. Initially, each observation was considered a separate cluster (i.e., a leaf in the dendogram). Then, the less distant clusters are successively merged until there is just one single cluster (i.e. the root of the dendogram). Several agglomeration methods have been proposed to determine the distance between clusters, such as the maximal distance between any two observations in the clusters, the average distance between the observations in the two clusters, the distance between centroids, or Ward's minimum variance criterion, which minimises the total within-cluster variance. 

\end{itemize}

The unsupervised nature of these methods makes them valuable for stakeholder selection,
as they try to discover patterns in data by assigning each observation to a previously unknown group. Clustering methods can be applied based on quantitative properties associated with stakeholders (i.e., salience). 
Once stakeholders are clustered, the remaining task is to identify (or perhaps relate somehow) a salience based category~\cite{mitchell1997} they fit in.

\section{Stakeholders categories according to salience} 

The salience model is the most popular and widely applied framework that we can use to categorise stakeholders, that is, to establish how much priority we should give a specific stakeholder~\cite{mitchell2019stakeholder}.
Salience is defined through the attributes \emph{power},  \emph{legitimacy}, and  \emph{urgency}~\cite{mitchell1997}. This theory produces a comprehensive typology of stakeholders supported by these three salience components, which are subjective in nature.

Power can be defined as the ability of a stakeholder to influence the outcome according to what they want. Power defines the potential capability to determine the actions to be performed by another stakeholder or a group of them. Legitimacy is the appropriateness that a stakeholder has from a socially defined set of norms and values point of view. Power and legitimacy are different, but they are coupled dimensions, and each can exist without the other. 

Urgency is a dynamic, time-related dimension that can be defined as the degree to which a stakeholder claims immediate attention. This framework considers the dynamic nature of three dimensions; any change in their value strongly affects the stakeholder priority in the system. 

These three attributes define the concept of stakeholder salience~\cite{mitchell1997, freeman1984}. In addition, according to their presence or absence, four different stakeholders' categories are defined:

\begin{itemize}
\item No stakeholders. They have no salience because none of the attributes is present.
\item Latent stakeholders. Their salience is low since only one attribute can be assigned to them.
\item Expectant stakeholders. They lack only an attribute that reaches moderate salience.
\item Definitive stakeholders. The three attributes are perceived by the managers to be present, so they have high salience.
\end{itemize}

A powerful stakeholder that has no legitimacy nor urgency keeps his power unused, with little interaction with the system. But if any change appears in the rest of the attributes, the stakeholder could change their category. When the stakeholder has only legitimacy, there is no pressure on the project managers to engage them in the system. The demanding stakeholder, those with only urgent claims with no power nor legitimacy, could be considered as latent noise that may become a real sound when some of the other attributes appear.

When power and legitimacy are present, stakeholder influence is taken for granted because they can decide about the system and are considered appropriate to define the system properties. Lacking power but having urgency and legitimacy is the case for dependent stakeholders  
that managers should take into account when defining the system. Additionally, the lack of legitimacy in a powerful stakeholder that makes urgent claims defines a dangerous stakeholder that could coercively influence the system.

Managers must respond to urgent claims proposed by powerful stakeholders that have legitimate interests according to social sets of norms and values. That is, stakeholders of high and moderate importance (that is, with at least two components of importance) must be taken into account~\cite{mitchell2019stakeholder}.  

Based on this stakeholder salience concept,  some other stakeholder identification methodologies propose five levels or categories taking into account the stakeholder attributes: power, legitimacy, knowledge, interest, and alliance with others~\cite{rahman2015}. A set of rules such as "if legitimacy is yes and the value is high for two variables and medium for the rest, the stakeholder has a good impact on the system", belonging to level 3. These rules allow managers to define the influence of each stakeholder.

The literature agrees that salience is the key point to know how many stakeholder groups can be found, but the number of groups varies; even classical requirements engineering works define only three stakeholder categories (critical, major, minor)~\cite{glinz2007guest}. In addition to salience components, estimations are usually qualitative or interval-based scores, such as lower range, mean, or upper range~\cite{rahman2015, poplawska2015}, making salience difficult to assess or estimate quantitatively.

Identifying inappropriate stakeholders (without salience) will lead to capture requirements that are not relevant to the real needs of the system. That is the reason that justifies the strong connection between requirement selection and the identification and quantification based on stakeholder theory.

\section{Materials and Methods}

Let $\mathbf{Stk}=\{sk_1,..sk_m\}$ be the set of $m$ stakeholders for a given project; our proposal is to automatically identify the most important stakeholders, $\mathbf{Def}\subset \mathbf{Stk}$, which should simplify the subsequent \ensuremath{\mathsf{NRP}} solving process that starts from the set $\mathbf{R}=\{r_1,..r_n\}$.
This set represents the candidate requirements to be included in the next software release.
To demonstrate how the proposed method can be adopted, we use a well-known data set that allows us to define the salience of stakeholders and at the same time be able to work with an \ensuremath{\mathsf{NRP}} instance.

\begin{figure}
   \centering
    \includegraphics[width=0.6\textwidth]{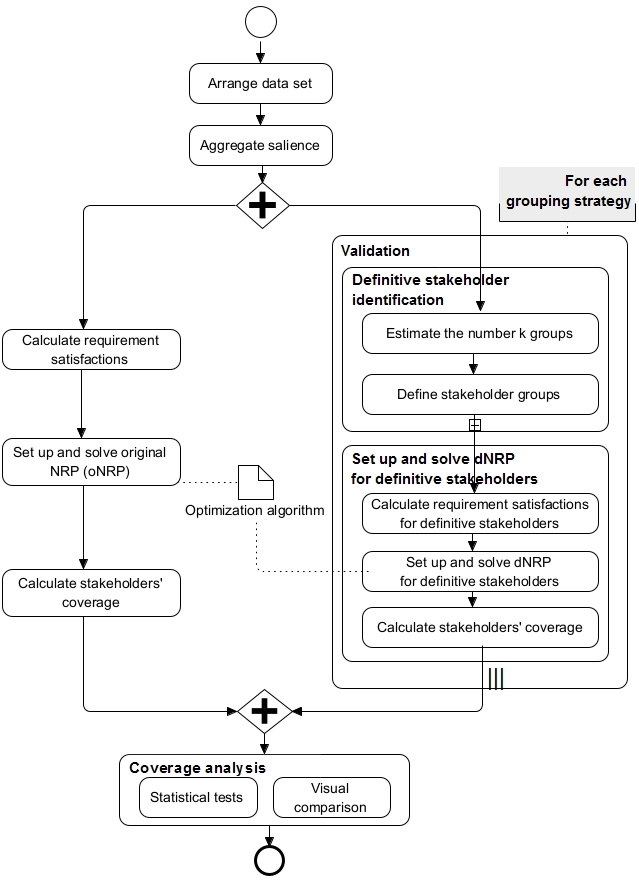}
   \caption{Methodology used as visual task flow.}

   \label{fig:workflow}
\end{figure}

The methodology applied to carry out our research is described using a visual task flow in figure~\ref{fig:workflow}. We start from a set of data on stakeholders and requirements. From the stakeholder data, we obtain the values of power, legitimacy, and urgency that are used to obtain the salience of each stakeholder (RQ1). Then, the satisfaction values associated with each of the requirements are obtained by adding the values suggested by the stakeholders, taking into account the importance each stakeholder has in the project. All these tasks result in the definition of the original requirement selection problem. The resolution of this problem will be carried out by applying an optimisation algorithm. The stakeholder coverage in the original problem will serve as a benchmark against which the stakeholder coverage results will be validated when we identify the definitive ones by applying grouping methods based on the values of the components that define the salience (RQ3).

\emph{For each grouping strategy}, an instance of a two-stage process is created:
\emph{definitive stakeholder identification} and \emph{set up and solve the requirements selection problem for definitive stakeholders}. The first stage estimates the number of stakeholder groups to consider, either following the number of categories suggested by Mitchell et al. \cite{mitchell1997}, or estimating it based on indices to establish the optimal number of clusters in a partition of a data set. Once the stakeholder groups have been obtained, the one with the highest salience is selected as the definitive stakeholder group (RQ2). Then a requirements selection problem is set up where the satisfaction values associated with each of the requirements are obtained by adding the values suggested by the definitive stakeholders taking into account the importance of the stakeholders in the project (i.e. \ensuremath{\mathsf{dNRP}}). This selection problem will be solved by applying the same optimisation algorithm as used in the original problem. 

Finally, \emph{coverage analysis} is carried out to check whether the results obtained considering only the final stakeholder group are sufficient to uniformly cover all initial stakeholders (considering as a measure of coverage the percentage of requested requirements that are included in the selected set of requirements). Thus, the stakeholders' coverage provided by each grouping strategy have to be analysed and compared to the benchmark (RQ3),  using both a statistical and a visual approach.

\subsection{Replacement Access, Library and ID Card project data set}

The data set used to study the research questions is \emph{Replacement Access, Library and ID Card project} (RALIC).
It was a software project to enhance the existing access control system at University College London (UCL).
The project combined different UCL access control
mechanisms into one, such as access to the library and fitness
centre, eliminating the need for a separate library registration process for UCL ID card holders~\cite{lim2011}.

It is a widely studied data set within the domain of Requirements Engineering that has been used in several works with many different approaches, such as 
the use of exact methods in \ensuremath{\mathsf{NRP}}~\cite{VEERAPEN20151}, the study of requirements interactions~\cite{ZHANG2013},  clustering technique for prioritization~\cite{achimugu2014} or the extraction from the RALIC data set of the values to be included in a Bayesian network that predicts the stability of a software requirements specification~\cite{sagrado2018}. The importance of stakeholders in RALIC data set has also been studied using genetic algorithms~\cite{lim2013} or clustering~\cite{zanaty2021}.

RALIC identifies stakeholders by creating a recommendation network. Each recommender selects a set of other stakeholders, giving them an influence level on the project. There are two defined networks, OpenR and ClosedR. The network we have selected is OpenR, because it includes more recomendees, 61, and more recommendations, 1714, in the range 1-8 each one. 
RALIC project involves 144 stakeholders in the networks, some of them acting only as a recommendee. However, not all recommendees or recommenders have put forward an enhancement or new functionality to include in the software to be built.

The collector of this data set also identifies 127 roles in the project, each with a priority according to the department to which they belong. Additionally, as an assignment to a role, stakeholders are ranked within the department / role. That is, the ground truth that established how powerful a stakeholder is according to their status in the project has been published~\cite{limthesis}. Similarly to the recommendation network, not all of them propose amendments for the project.

The project includes 138 requirements as increases to
the actual access, library and ID card system, which are arranged in three levels: objectives (10), requirements (48), and specific requirements (104), they are represented by $\mathbf{R}=\{r_1,r_2, \ldots ,r_n\}$. Their effort range varies from 4 to 7000 persons-hour. Only 75 RALIC stakeholders use the 100-point method (each stakeholder is given 100 points that can be used to vote for the most important requirements) to prioritise the requirements they are interested in; $\text{points}_{ij}$ represents the votes that the stakeholder $i$ assigns to the requirement $r_j$ that can be used to calculate requirement satisfaction. However, stakeholders do not vote at the same level of the three defined ones.
These facts translated into the reorganisation of requirements (e.g. requirements that nobody asks for are erased), obtaining for our study 83 requirements,  $\mathbf{R}=\{r_1,r_2, \ldots ,r_{83}\}$ and their corresponding development efforts in $\mathbf{E}=\{e_1,e_2, \ldots ,e_{83}\}$.

\subsection{Arranging stakeholders based on salience distribution} 
\label{sec:salience_select}

Power ($\text{power}_i$), legitimacy ($\text{legitimacy}_i$) and urgency ($\text{urgency}_i$) of a stakeholder ($sk_i$) have been extracted from the RALIC data set using, respectively, the ground truth for stakeholder status assigned to $sk_i$,  the total of recommendations from the OpenR network received by $sk_i$, and the sum of points assigned to all requirements by $sk_i$. In this way, we define the salience of $sk_i$ as follows:

\begin{equation}
\label{eq:salience}
    \text{salience}_i = \text{power}_i + \text{legitimacy}_i + \text{urgency}_i, \quad \forall sk_i \in \mathbf{Stk}.
\end{equation}

Ordering RALIC stakeholders by salience is trivial without taking into account that it has to be assessed from the value of its three components according to stakeholders theory because a very powerful stakeholder that has not been legitimated by their project mates, even having demands (urgency), could be considered less salient than the one with a medium value assigned to the three salience components.

\begin{figure}
   \centering
    \includegraphics[width=0.8\textwidth]{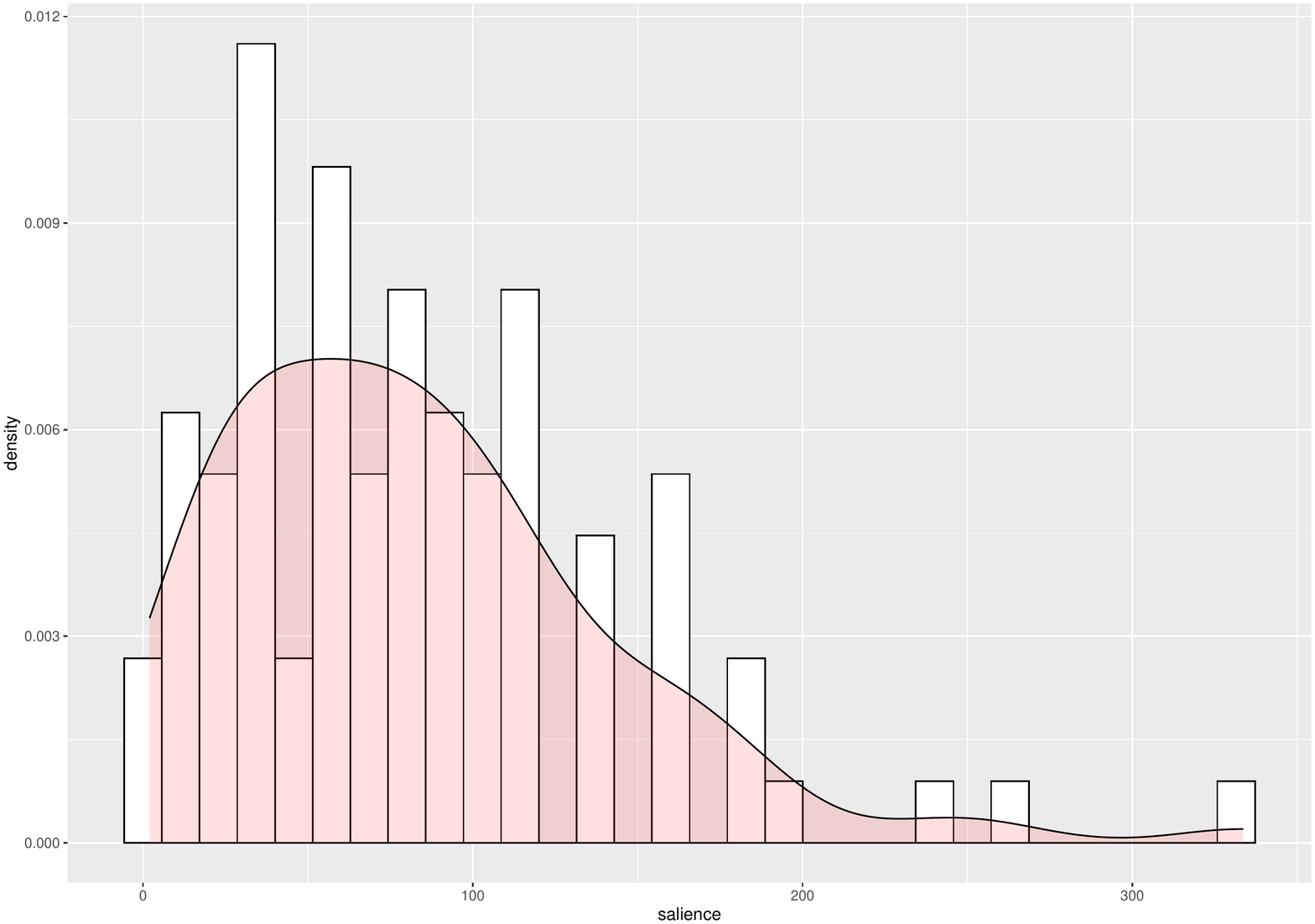}
   \caption{Histogram and density plot of definitive and expectant stakeholders' salience.}

   \label{fig:salience}
\end{figure}

The case to study has a set of stakeholders $\mathbf{Stk}$ with 98 definitive and expectant stakeholders (because 46 only have one salience component), and a set of requirements $\mathbf{R}$ with 83 requirements ($m=98$, $n=83$, for the sets $\mathbf{Stk}$ and $\mathbf{R}$). Figure \ref{fig:salience} shows the salience distribution and Table \ref{tab:salience} contains the statistical summary of salience, power, legitimacy, and urgency for the definitive and expectant stakeholders. Salience can be considered as the first approach to weight stakeholders in \ensuremath{\mathsf{NRP}}.

\begin{table}
\caption{Statistic summary of salience components for the definitive and expectant stakeholders}
\label{tab:salience}
\begin{tabular}{lrrrrrr}
    \hline
 &
  \multicolumn{1}{l}{\textbf{Min.}} &
  \multicolumn{1}{l}{\textbf{1st Qu.}} &
  \multicolumn{1}{l}{\textbf{Median}} &
  \multicolumn{1}{l}{\textbf{Mean}} &
  \multicolumn{1}{l}{\textbf{3rd Qu.}} &
  \multicolumn{1}{l}{\textbf{Max.}} \\ \hline
\textit{power}      & 0.00 & 0.00  & 19.50 & 19.92 & 36.50  & 49.50  \\
\textit{legitimacy} & 0.00 & 12.25 & 33.50 & 41.62 & 63.75  & 251.00 \\
\textit{urgency}    & 0.00 & 3.75  & 20.00 & 21.52 & 34.75  & 72.00  \\
\textit{salience}   & 2.00 & 37.00 & 73.00 & 83.07 & 113.05 & 333.50 \\   \hline
\end{tabular}
\end{table}

When shifting to the \ensuremath{\mathsf{NRP}} domain,  the overall revenue or satisfaction of a given
requirement has to be calculated according to the combination of the values estimated by each stakeholder as requirement benefit. Most of the already validated proposals use the weighted sum of values, which is called satisfaction. 
The satisfaction of a requirement $j$ given a set of stakeholders $\mathbf{D} \subseteq \mathbf{Stk}$, namely $\text{points}_{ij}$ the votes that the stakeholder $i$ assigns to the requirement $r_j$ in RALIC, can be calculated as

\begin{equation} \label{eq:sat}
    \text{sat}_j = \sum_{i \in \mathbf{D}} \text{salience}_i *\text{points}_{ij}.
\end{equation}

From these three sets, $\mathbf{Stk}$, $\mathbf{R}$, and $\mathbf{E}$, together with the definition of the requirements satisfaction values (equation \ref{eq:sat}), we can define the overall next release problem for RALIC (including all stakeholders), \ensuremath{\mathsf{oNRP}}, as the following optimisation problem:

\begin{equation}\label{eq:optimize}
\begin{array}{l}
   \max \sum_{j \in \mathbf{U}} \text{sat}_j, \\

   \min \sum_{j \in \mathbf{U}} e_j,\\
   \text{subject to }  \qquad    B_1 \leq \sum_{j \in \mathbf{U}} e_j \leq B_2.

\end{array}
\end{equation}

 \noindent where $\mathbf{U}$ is a solution (i.e., a set of requirements that conforms to the next release).  The limits of resources / effort are defined in the range $[B_1$, $B_2]$ which, respectively, corresponds to the 20 \% the effort needed to develop all requirements for $B_1$, while $B_2$ is 25 \%. These values have been chosen taking into account the contingency value for effort, that is, an allowance made for the risk that something will not be undertaken with the planned estimate effort. We have decided to define a resource limit interval because, on the one hand, the original data set did not include an upper resource limit and, on the other hand, developers usually discard solutions in the Pareto front with low effort value~\cite{aguila2016}. For RALIC the values for $B_1$ and $B_2$ are, respectively, 12473.3 and 13304.8.

 \ensuremath{\mathsf{NRP}} is the problem we want to downsize, reducing the number of stakeholders to take into account. Initially, we computed the salience values (see equation \ref{eq:salience})  taking into account only definite and expectant stakeholders, because according to stakeholder theory, a stakeholder to be considered should at least have two salience components. Expectant stakeholders category is divided into three subcategories, corresponding to dominant, dependent, and dangerous stakeholders, respectively. Thus, we need to consider four categories when determining which stakeholders are important~\cite{mitchell1997}.

Summary statistics communicate the largest amount of information, as simply as possible, about the distribution of the values of a variable. And a quartile is any of the three values that divide the ordered set of data into four equal parts, so each part represents $1/4$ of the sample or population. Each part could roughly be considered as one of the stakeholder groups. Specifically, the third quartile is the value above which 25\% of the highest values are found.

With this initial idea in mind, based on the salience distribution (Figure \ref{fig:salience} and Table \ref{tab:salience}), the 25\% of stakeholders that have the highest salience value (that is, greater than the third quartile) could be simplistically considered as the group of the most influential and identified as the important stakeholders (definitive) in terms of salience.  In addition to this statistically based arrangement strategy, there are other techniques to make groups such that stakeholders in the same group were more similar to each other than to those in the other group (i.e., clustering), but considering the salience components.

\subsection{Finding stakeholders groups}
\label{sec:clustering_select}

We propose to search for stakeholder groups using the values of power, legitimacy, and urgency as data input for clustering. The most common clustering techniques usually fall into two categories depending on whether they need to know the number of clusters to be searched in advance (i.e. \emph{partitioning methods}, such as \emph{k-means}~\cite{hartigan1979} and \emph{k-medoids}~\cite{kaufman1990.ch2}) or not (i.e. \emph{hierarchical methods}, such as \emph{agglomerative clustering}~\cite{murtagh2012}). 
Within partitioning algorithms \emph{k-means} is simple, fast, and can deal with large data sets, but it is sensitive to data ordering and anomalous data points and outliers. While being more reliable and less sensitive to outliers, \emph{Partitioning around medioids}(\emph{k-medoids}) requires more computational effort than \emph{k-means}. In contrast to them, \emph{agglomerative hierarchical clustering} groups observations on the basis of their similarity. Initially, each observation is considered a separate cluster (i.e., a leaf in the dendogram). Then, the less distant clusters are successively merged until there is just one single cluster (i.e. the root of the dendogram).

These methods, which are unsupervised in nature, attempt to discover patterns in the data by assigning each observation (i.e., a stakeholder's values of power, urgency, and legitimacy) to a group that is not known beforehand. Once the stakeholder groups have been discovered, it will be necessary to identify which are the important or definitive ones.

To apply the clustering algorithms suggested above, it should be noted that the number $k$ of stakeholder groups can be set to $4$, coinciding with the categories (i.e., definitive, dominant, dependent and dangerous, the last three are subclasses of expectant) suggested by Mitchell et al.~\cite{mitchell1997}. Or, we can resort to a variety of proposed indices to find the optimal number of clusters in a partition of a data set~\cite{Charrad2014}.
Deciding on the number of clusters that suitably fit a data set is a problem itself. It involves simultaneously evaluating the clustering method using validation indices while varying the number of clusters. The \emph{R} package \emph{NbClust}~\cite{Charrad2014}, given a clustering method and a data set, obtains the best number of groups for a set of 30 validation indices by applying the majority rule to the best number of clusters returned by each index. That is, the best number of clusters is the one in which the largest number of indices coincide.
Then, we have to check if the expected relationship between the groups of stakeholders discovered by clustering and Mitchell's stakeholder categories occurs.

\subsection{Validation }
\label{subsec:workflow}

We have defined a set of tasks to follow to provide an answer to each posed research question and to demonstrate the applicability of reducing the number of stakeholders (making groups) to be involved in a requirement selection process that keeps stakeholders covered (see figure \ref{fig:workflow}):

\begin{itemize}

\item[i)] \emph{Definitive stakeholders identification}. This process has as goal determining the stakeholders that really count for the project. We can proceed in two ways: based on salience or on clustering methods. If the identification is based on salience (see Section \ref{sec:salience_select}), the definitive stakeholders identified would be those whose salience is greater than salience Q3. Whereas if we use clustering-based selection (see Section \ref{sec:clustering_select}), we will have to set the number of clusters, use a clustering method, find the clusters, and decide which of the clusters corresponds to that of definitive stakeholders. 

Once the clusters are found, the group with definitive stakeholders 
($\mathbf{Def}\subset \mathbf{Stk}$) has yet to be identified. The rule we use is to select the cluster of stakeholders that achieves the highest average value in power, legitimacy, and urgency, as the definitive stakeholders are those with higher salience.

Each clustering algorithm is expected to identify different stakeholder groups and, consequently, the definitive stakeholders will not match. The same will happen when we compare them with the final stakeholders identified using only quartiles for the salience values.

\item[ii)] \emph{Setup and solve an \ensuremath{\mathsf{NRP}} for definitive stakeholders}.
Once the set $\mathbf{Def}\subset \mathbf{Stk}$ of definitive stakeholders has been established, the initial \ensuremath{\mathsf{oNRP}} (see equation \ref{eq:optimize}) will be altered because the set of stakeholders is now that of the definitive ones. Consequently, the satisfaction of each requirement $j \in \mathbf{R}$ has to be recalculated by applying the equation \ref{eq:salience}, but only using the salience and points values of this subset of stakeholders. We refer to the new optimisation problem as \ensuremath{\mathsf{dNRP}}. After solving it (using any of the algorithms proposed in the literature~\cite{nuh2021}) we get a set of non-dominated solutions defining a Pareto front $\mathbf{F_{Def}}$.

Since studying the impact of each of the selected definitive stakeholder groups, several requirements selection problems \ensuremath{\mathsf{dNRP}} need to be defined based on each definitive stakeholder group. It is necessary to apply the same algorithm to find solutions to each of the optimisation problems. In this way, the solutions will be comparable.

\item[iii)] \emph{Stakeholders coverage analysis}. 
For a given \ensuremath{\mathsf{NRP}} solution, $\mathbf{U}$, we defined a measure of how this solution covers the desires of a stakeholder $i \in \mathbf{Stk}$~\cite{aguila2016}. Since the 100-points method is applied in stakeholders' estimations, we express can this metric as

\begin{equation}\label{eq:stcov}
    \text{stcoverage}_i(\mathbf{U})=  {\sum_{j \in {\mathbf{U}}}\text{points}_{ij}}/ 100.
\end{equation}

\noindent Based on this function, we can express the coverage of a stakeholder $i \in \mathbf{Stk}$ provided by the Pareto front $\mathbf{F_{Def}}$, obtained as a solution of a \ensuremath{\mathsf{dNRP}}, when only the set of definitive stakeholders ($\mathbf{Def} \subset \mathbf{Stk}$) is considered,  as the arithmetic mean of coverage for all its solutions. 

\begin{equation}\label{eq:cov}
    \text{cov}_i (\mathbf{F_{Def}})=  \frac{1}{\#\mathbf{F_{Def}}} \sum_{\mathbf{U} \in \mathbf{F_{Def}}}
     \text{stcoverage}_i(\mathbf{U}).
\end{equation}

This measure serves as a basis for assessing the effectiveness of the stakeholder selection we propose. The best selection strategy will be the one that provides the most coverage and to the most initial stakeholders ($\mathbf{Stk}$).

We will compare the coverage of the solutions obtained for each of the \ensuremath{\mathsf{dNRP}}s against the coverage obtained in the \ensuremath{\mathsf{oNRP}} without selecting stakeholders 
and the \ensuremath{\mathsf{dNRP}}s with each other, selecting as the definitive stakeholder group the one that induces greater coverage of all stakeholders.

\end{itemize}

We applied these steps in the study to the RALIC data set. First, we obtain several definitive stakeholders groups, using several identification technique (that is, grouping strategies). Next, we searched for the solutions of each of the \ensuremath{\mathsf{dNRP}}s using the same optimisation algorithm. Finally,
we have studied the stakeholder coverage for $\mathbf{Stk}$  that each solution provides. The comparison of the solutions obtained is made using as a benchmark the solutions obtained for \ensuremath{\mathsf{oNRP}} considering all stakeholders in $\mathbf{Stk}$ and the same optimisation algorithm  (see figure \ref{fig:workflow}).
All experiments were programmed in R and \ensuremath{\mathsf{NRP}} instances were solved using a 2,6 GHz CPU machine, with two kernels and 8 GB of RAM.

\section{Results and discussion}

\subsection{Definitive stakeholders identification}

\begin{figure}
   \centering
    \includegraphics[width=0.5\textwidth]{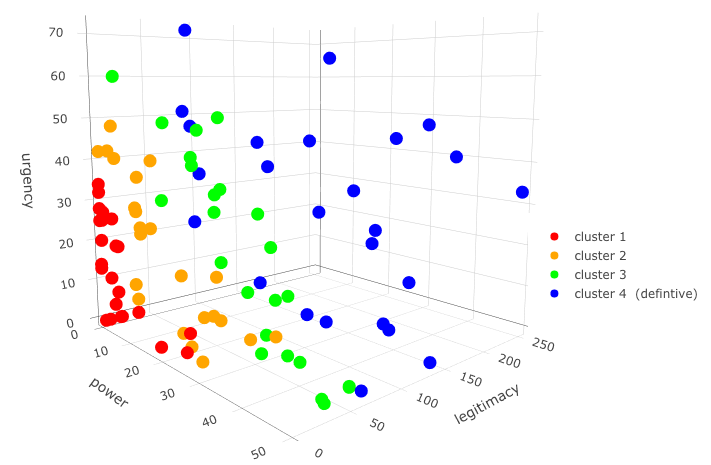}
   \caption{Stakeholders groups defined based on salience quartiles.}
   \label{fig:quartiles_clusters}
\end{figure}

As the first grouping strategy  
to identify the definitive stakeholders in the RALIC data set, we can use the salience values and locate the 25\% of stakeholders whose salience is above the third quartile (i.e., a salience greater than 113.05). We denote by ($\mathbf{Def_1}$) the group of stakeholders that satisfies this condition. In addition, we can similarly use the salience quartiles to divide stakeholders into other groups: those whose salience is less than or equal to the first quartile, those whose salience is greater than the first quartile but less than or equal to the second quartile, and those whose salience is greater than the second quartile but less than or equal to the third quartile.  Table \ref{tab:Q3-salience} shows the number and average values of power legitimacy and urgency of the four stakeholder groups just defined. Figure \ref{fig:quartiles_clusters} depicts the distribution of the groups in terms of salience components, and it can be seen that there is no clear separation between them. Due to this fact, it is worthwhile to study other 
strategies to  obtain stakeholder groups, such as k-means, k-medoids and hierarchical clustering.

\begin{table}
\caption{Size and average values of the salience components of the stakeholder group whose salience exceeds the third quartile.}
\label{tab:Q3-salience}
\begin{tabular}{lcrrrc}
\hline
\textit{\textbf{cluster ID}} &
\textit{\textbf{size}} & \textit{\textbf{power}} & \textit{\textbf{legitimacy}} & \textit{\textbf{urgency}} & \textit{\textbf{definitive stk.}}\\ \hline
4 ($> Q3$) & 25 & 34.91 & 94.40 & 31.88 &$\mathbf{Def_1}$\\
3 $(Q2, Q3]$ & 24 & 30.06 & 38.83 & 23.17 &\\
2 $(Q1, Q2]$ & 23 & 12.27 & 25.35 & 19.61 &\\
1 ($\leq Q1]$ & 26 & 2.93 & 7.85 & 11.73 &\\
\hline
\end{tabular}
\end{table}

\begin{figure}
   \centering
    \includegraphics[width=0.5\textwidth]{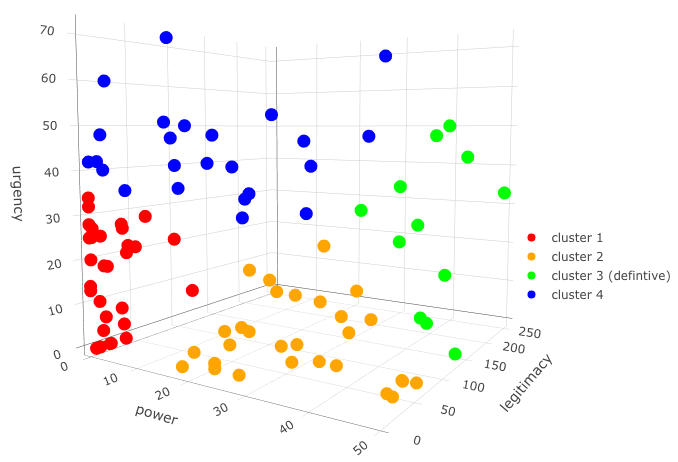}
   \caption{Stakeholders groups found by k-means and k-medoids when the number of clusters is set to 4.}
   \label{fig:k-means_pam_4_clusters}
\end{figure}

\begin{figure}
   \centering
    \includegraphics[width=0.5\textwidth]{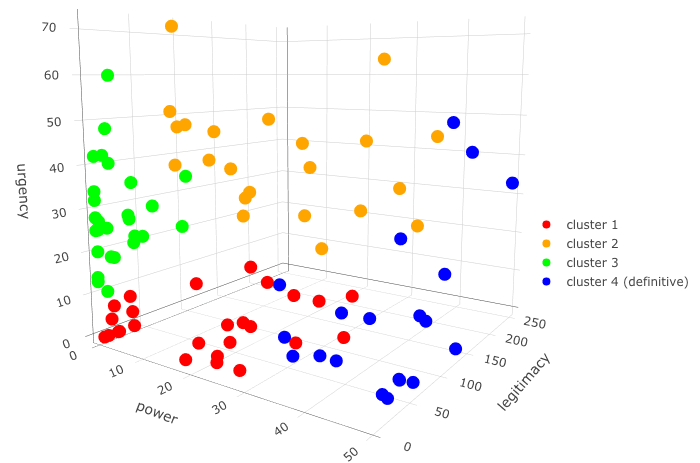}
   \caption{Stakeholders groups found by hierarchical clustering when the number of clusters is set to 4.}

   \label{fig:hc_4_clusters}
\end{figure}

With k-means, according to the majority rule, the best number of clusters is 4. The next step is to obtain 4 stakeholder groups by applying k-means, k-medoids, and hierarchical clustering. Figures \ref{fig:k-means_pam_4_clusters} and \ref{fig:hc_4_clusters} show the clusters found. Then, we will calculate the centroids (i.e. a reference point that is the average of the points forming the cluster) of the groups obtained with each of the techniques and identify the definitive stakeholder group as the one whose centroid reaches the highest salience value. Table \ref{tab:4K-clusters} shows the clusters found by the different clustering methods characterised by their size and centroid. The centroids are ordered by salience value in decreasing order, so for each method, we will take the first group as definitive stakeholders (i.e., cluster 3 for k-means and k-medoids, and cluster 4 for hierarchical clustering). 

\begin{table}
\caption{Centroids of the stakeholder groups returned by the clustering methods using 4 clusters.}
\label{tab:4K-clusters}
\begin{tabular}{cccrrrc}
\hline
\textit{\textbf{method}} & \textit{\textbf{cluster ID}} & \textit{\textbf{size}} & \textit{\textbf{power}} & \textit{\textbf{legitimacy}} & \textit{\textbf{urgency}} & \textit{\textbf{definitive stk.}} \\ \hline
                      & 3 & 12 & 46.78 & 120.25 & 28.17& $\mathbf{Def_2}$  \\
 \textit{k-means}      & 2 & 29 & 33.14 & 39.28  & 6.97  \\
 \textit{k-medoids}    & 4 & 24 & 16.88 & 36.00  & 45.08 \\
                      & 1 & 33 & 0.76  & 19.18  & 14.76 \\ \hline
                      & 4 & 20 & 42.34 & 93.25  & 11.15 & $\mathbf{Def_3}$ \\
\textit{hierarchical} & 2 & 22 & 8.32  & 46.73  & 43.50 \\
                      & 1 & 29 & 16.52 & 19.86  & 6.00  \\
                      & 3 & 27 & 0.14  & 22.59  & 27.96 \\ \hline
\end{tabular}
\end{table}

\begin{figure}
   \centering
    \includegraphics[width=0.5\textwidth]{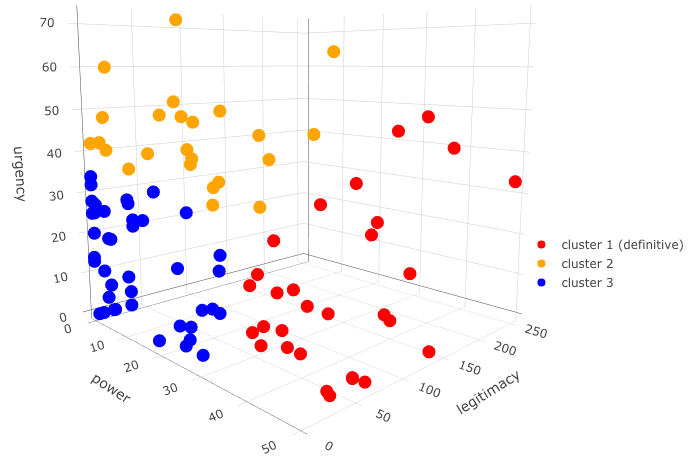}
   \caption{Stakeholders groups found by k-means when the number of clusters is set to 3.}

   \label{fig:k-means_3_clusters}
\end{figure}

\begin{figure}[ht]
   \centering
    \includegraphics[width=0.5\textwidth]{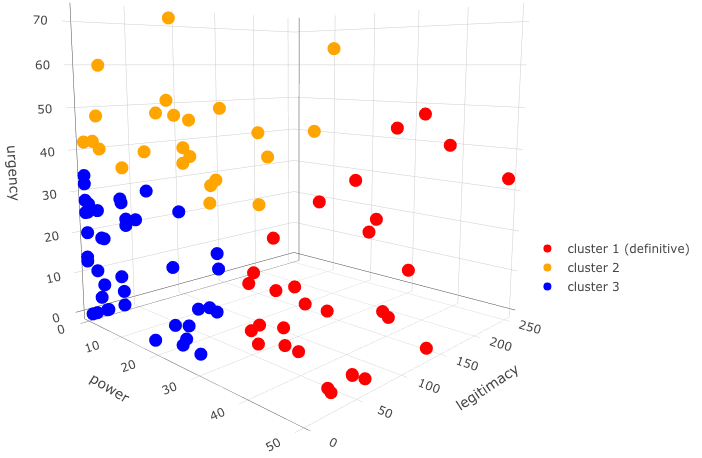}
   \caption{Stakeholders groups found by k-medoids when the number of clusters is set to 3.}

   \label{fig:pam_3_clusters}
\end{figure}

\begin{figure}
   \centering
    \includegraphics[width=0.5\textwidth]{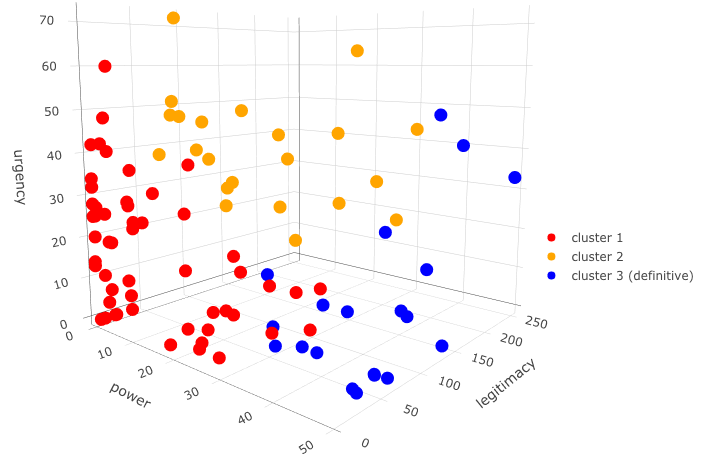}
   \caption{Stakeholders groups found by hierarchical clustering when the number of clusters is set to 3.}
   \label{fig:hc_3_clusters}
\end{figure}

Analogously, when we use k-medoids, according to the majority rule, the best number of clusters is 3. Now, we apply the different clustering methods to obtain 3 stakeholder groups. Figures \ref{fig:k-means_3_clusters}, \ref{fig:pam_3_clusters}, and \ref{fig:hc_3_clusters} show the stakeholder groups identified. Following the previous considerations, Table \ref{tab:3K-clusters}  collects the sizes and centroids of the clusters found. For each method, the first cluster in the table is selected as the definitive stakeholder group, since it has the highest salience (i.e., cluster 1 for k-means and k-medoids, and cluster 3 for hierarchical clustering). In the case of hierarchical clustering, regardless of whether the number of groups is set to 3 or 4, the definitive set of stakeholders is the same $\mathbf{Def_3}$.

\begin{table}
\caption{Centroids of the stakeholder groups returned by the clustering methods using 3 clusters.}
\label{tab:3K-clusters}
\begin{tabular}{cccrrrc}
\hline
\textit{\textbf{method}} & \textit{\textbf{cluster ID}} & \textit{\textbf{size}} & \textit{\textbf{power}} & \textit{\textbf{legitimacy}} & \textit{\textbf{urgency}} & \textit{\textbf{definitive stk.}}\\ \hline
                      & 1 & 30 & 42.06 & 78.30 & 15.70   & $\mathbf{Def_4}$\\
\textit{k-means}      & 2 & 24 & 16.88 & 36.00 & 45.08  \\
                      & 3 & 44 & 6.50  & 19.68 & 12.64\\ \hline
                      & 1 & 33 & 40.53 & 74.27 & 15.33 & \textbf{$\mathbf{Def_5}$}\\
\textit{k-medoids}    & 3 & 28 & 15.67 & 36.64 & 42.36 \\
                      & 2 & 37 & 4.77  & 16.27 & 11.27 \\ \hline
                      & 3 & 20 & 42.34 & 93.25 & 11.15  & $\mathbf{Def_3}$ \\
\textit{hierarchical} & 2 & 22 & 28.32 & 46.73 & 43.50 \\
                      & 1 & 56 & 8.62  & 21.18 & 16.59 \\ \hline
\end{tabular}\\
\end{table}

The results reported in this subsection provide a positive answer to the RQ1 and RQ2 questions. The definition of stakeholder salience proposed in equation \ref{eq:salience} allows a quantitative definition of stakeholder importance which also serves as a basis to identify the most important ones (i.e., definitive stakeholders) for \ensuremath{\mathsf{NRP}} (see Fig. \ref{fig:quartiles_clusters}). In this way, the first goal is achieved: The weight of stakeholders in the requirement selection problem is matched by the concept of salience. 
Besides, clustering techniques define an appropriate number of stakeholders groups compatible with stakeholders salience theory (see Figs. \ref{fig:k-means_pam_4_clusters}  to \ref{fig:hc_3_clusters}), which reinforces the positive answer to RQ2. 
The clusters found define separate stakeholder groups according to power, legitimacy, and urgency. However, five different cluster configurations are compatible with stakeholder salience theory.   Independently of the cluster configuration used, the group of stakeholders identified as the definitive stakeholder group is the group with the highest importance, allowing automatic identification of the most important stakeholders (RQ1).  What remains to be verified to fulfil our goal is if a uniform covering of all stakeholders implied in the requirements selection problem can be obtained by considering only the group of definitive stakeholders (RQ3). The following sections deal with this issue.

\subsection{Set up and solve an  \ensuremath{\mathsf{NRP}} for each definitive stakeholder group}

After identifying different sets of definitive stakeholders (i.e.,$\mathbf{Def_1}$ to $\mathbf{Def_5}$) depending on the method and number of groups used, the next  task  is to define and solve the corresponding \ensuremath{\mathsf{dNRP}} for each of these sets (see task \emph{ii} in Section \ref{subsec:workflow} and figure \ref{fig:workflow}). To find the Pareto front (i.e. $\mathbf{F_{Def_1}}$ to $\mathbf{F_{Def_5}}$) for each of these problems, we have resorted to a 
greedy algorithm. This algorithm works as follows: for each effort value in the range $[B1,B2]$ we fill the solution with as many requirements as we can. Next, within a fixed number of attempts, we try to remove one by one each requirement in the solution, replacing it with another that is valid for that level of effort. After every change, the dominance is checked, substituting the solution if needed.

Although it may not be the best algorithm to find the Pareto fronts (which is beyond the scope of this paper~\cite{nuh2021, zhang2018}), it is at least simple 
and obtains the Pareto fronts in the same way in all cases, avoiding biases that would occur in experiments if different methods were used to obtain solutions.

In addition to the above cases, we will also consider the group where all stakeholders are taken into account (i.e., $\mathbf{Def_0} \subseteq \mathbf{Stk}$), since it defines the \ensuremath{\mathsf{oNRP}} problem (see equation \ref{eq:optimize}) and we will solve it by applying the same 
greedy algorithm to obtain its Pareto front $\mathbf{F_{Def_0}}$. The coverage provided by this Pareto front serves as a benchmark to assess the level of coverage achieved with the different selected sets of definitive stakeholders.

\subsection{Stakeholders coverage analysis}
In the case of RALIC, only 75 of the 98 stakeholders in $\mathbf{Stk}$ are demanding, $\mathbf{Dem} \subset \mathbf{Stk}$, that is, they demand improvements for the next software release. The coverage analysis is based on the study of coverage (see equation \ref{eq:cov}) of these demanding stakeholders provided by the different Pareto fronts returned before and after selecting the definitive stakeholders.
The second column (that is, \emph{demanding / total}) in Table \ref{tab:summary} includes the number of demanding stakeholders versus the total, both in the initial stakeholder set ($\mathbf{{Def_0}}$) and in the selected stakeholder groups. 
Thus, for each Pareto front $\mathbf{F_{Def_h}}, h \in \{0,1,2, \cdots, 5\}$, the coverage for each demanding stakeholder, $cov_i(\mathbf{F_{Def_h}})$, $i \in \mathbf{Dem} \subset \mathbf{Stk}$, is calculated. First, we quantitatively analyse the stakeholder coverage provided by the Pareto fronts from a statistical point of view.  Second, we graphically represent the coverage and discuss some specific situations discovered in the data set.

As a summary of stakeholder coverage on each front, we use its mean and standard deviation (see column \emph{Average. stk. cov.} of Table \ref{tab:summary}). The dispersion of the coverage values with respect to their mean (see column \emph{Coef. Var.} of Table \ref{tab:summary}) is measured by the coefficient of variation (i.e., standard    deviation divided by the average). It can be seen that with $\mathbf{Def_5}$ (i.e., k-medoids clustering using 3 groups) the same coverage dispersion is obtained as when no stakeholder group is selected ($\mathbf{Def_0}$), while with the other groups there is an increase in dispersion. The smallest increase, which is 50\%, is obtained with $\mathbf{Def_4}$ (that is, k-means clustering using 3 groups).

In terms of the number of stakeholders covered with respect to the coverage achieved using $\mathbf{Def_0}$ (see column \emph{Stk. coverage with respect to $\mathbf{F_{Def_0}}$} of Table \ref{tab:summary}), selections $\mathbf{Def_3}$ (i.e. hierarchical clustering) and $\mathbf{Def_4}$ (that is, k-means clustering using 3 groups) are the best performers. They give the same coverage as that obtained without any selection (i.e. $\mathbf{Def_0}$) for 18.67\% of the demanding stakeholders (14/75), higher for 57.33\% (43/75) and worse for 24\% of them (18/75). However, with $\mathbf{Def_4}$ slightly better coverage is obtained on average because definitive stakeholders demand a set of requirements that better matches each other's requests.

\begin{table}
\caption{Values for Coverage Analysis}
\label{tab:summary}
\centering
\begin{tabular}{cccccccccc}
\hline
Group  &	\emph{demanding/total} & Average&Coef.&\multicolumn{3}{c}{Stk. coverage with respect to $\mathbf{F_{Def_0}}$} \\ \cline{5-7}
       &                &  stk. cov.&Var.& stk with $<$ cov& stake with $=$ cov & stk with $>$ cov \\
\hline
 $\mathbf{Def_0}$  &	75/98  &   \emph{84.89}$\pm$ 17.29 &0.20 & 
 \\
 $\mathbf{Def_1}$  &    23/25  &   71.37 $\pm$24.27  & 0.34 & 65	&  4&   6  \\  
 $\mathbf{Def_2}$  &	11/12  &  79.38 $\pm$ 32.65 & 0.41 & 21	& 11&	43  \\
 $\mathbf{Def_3}$  &	10/20  &  80.62 $\pm$ 30.45 & 0.38 & 18	& 14&	43  \\
 $\mathbf{Def_4}$  &	20/30  &  83.90 $\pm$ 25.15 & 0.30 & 18	& 14&	43  \\
 $\mathbf{Def_5}$  &	23/33  & 85.72 $\pm$ 17.19 & 0.20 & 31	& 15&	29  \\
\hline
\end{tabular}

\end{table}

We use the Wilcoxon signed-rank test to compare the coverage that the Pareto front $\mathbf{F_{Def_0}}$ initially provides to the demanding stakeholders (i.e. the benchmark) with that provided by each of the fronts $\mathbf{F_{Def_h}}, h \in \{1,2,3, \cdots, 5\}$ obtained after determining the set of definitive stakeholders (see column $\mathbf{F_{Def_0}}$ of Table \ref{tab:wilcoxon}). In this way, we compare the mean rank of the two samples and determine if there are statistically significant differences between them. As can be seen, except for the case of selection of stakeholders based on the salience distribution quartiles (i.e. $\mathbf{F_{Def_1}}$ whose p-value is less than or equal to 0.05), there are no statistically significant differences between the coverage values when using all demanding stakeholders (i.e. $\mathbf{F_{Def_0}}$) and those obtained when selecting the final stakeholders using clustering techniques (i.e. $\mathbf{F_{Def_h}}$, $h \in \{2,3, \cdots, 5\} $). Furthermore, there are statistically significant differences between
the coverage provided by the quartile-based stakeholder selection 
and that provided by the clustering methods (see column $\mathbf{F_{Def_1}}$ of Table \ref{tab:wilcoxon}). 
Statistically significant differences are also detected in the coverage of stakeholders obtained using 
the definitive stakeholder sets $\mathbf{Def_4}$ and $\mathbf{Def_5}$ (that is, the p-value is less than 0.05 between $\mathbf{F_{Def_4}}$ and $\mathbf{F_{Def_5}}$ in Table \ref{tab:wilcoxon}). In this case, although the average coverage obtained in $\mathbf{F_{Def_5}}$ is higher, the number of stakeholders with coverage greater than or equal to that obtained in $\mathbf{F_{Def_0}}$ is lower (58.67\% compared to 76\% in $\mathbf{F_{Def_4}}$). However, there are no significant differences in the coverage values obtained when the definitive stakeholders are selected using clustering (see columns $\mathbf{F_{Def_2}}$ and $\mathbf{F_{Def_3}}$ of Table \ref{tab:wilcoxon}), regardless of the technique and number of groups (that is, 3 or 4) used.

\begin{table}
\caption{p-value of the Wilcoxon test for Pareto fronts $\mathbf{ F_{Def_h}}$}
\label{tab:wilcoxon}
\begin{tabular}{ccccccc}
\hline
   & $\mathbf{F_{Def_0}}$& $\mathbf{F_{Def_1}}$ & $\mathbf{F_{Def_2}}$ & $\mathbf{F_{Def_3}}$ & $\mathbf{F_{Def_4}}$  \\ \hline

$\mathbf{ F_{Def_1}}$ &  2.054e-10 \cellcolor{gray!15}  \\     
$\mathbf{ F_{Def_2}}$  & 0.623 &  7.63e-07 \cellcolor{gray!15} \\
$\mathbf{F_{Def_3}}$ & 0.417 & 1.496e-08  \cellcolor{gray!15}  & 0.1049\\
$\mathbf{F_{Def_4}}$ & 0.06809 & 7.352e-10  \cellcolor{gray!15}  & 0.639 & 0.5754\\ 
$\mathbf{F_{Def_5}}$ & 0.8453  & 1.156e-09  \cellcolor{gray!15} &  0.3457 & 0.3031&  0.03359 \cellcolor{gray!15}\\ \hline
\end{tabular}
\end{table}

\begin{figure}
   \centering
      \includegraphics[width=0.99\textwidth]{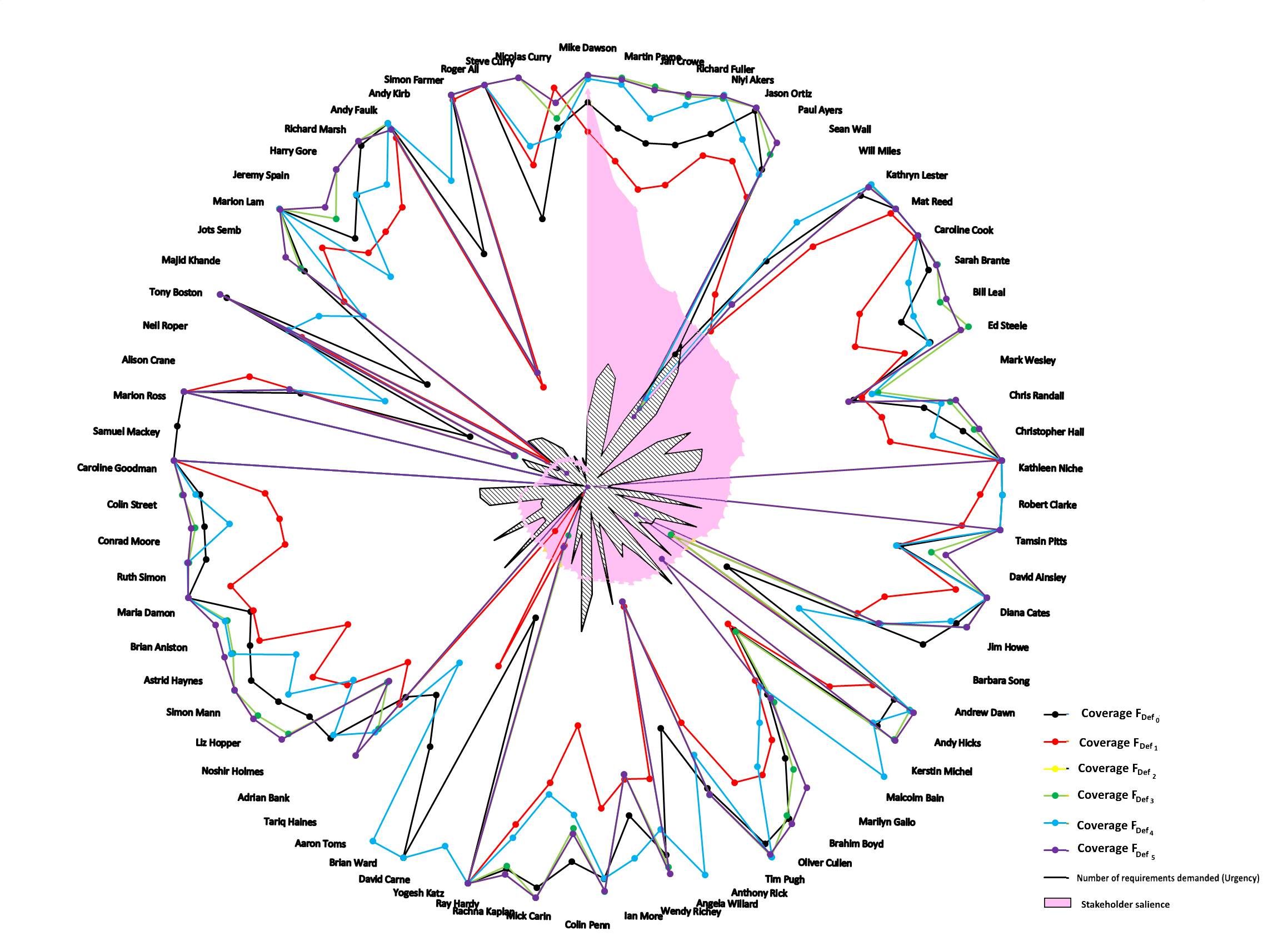}
   \caption{ Coverage provided by each Pareto front ($\mathbf{F_{Def_i}}$), salience (shaded in pink) and number of requirements demanded (shaded in grey) for the  75 demanding  stakeholders.}

   \label{fig:todos}
\end{figure}

Figure \ref{fig:todos} shows a Kiviat chart that includes the coverage of all stakeholders who demand requirements in RALIC, regardless of whether they are considered important or not. According to equations \ref{eq:stcov} and \ref{eq:cov} the axis range is given in the range from 0 to 100 points. The areas shaded in grey and pink represent the number of requirements demanded by each stakeholder and her/his salience, respectively. The stakeholders have been arranged clockwise in the graph, from highest to lowest salience, and the area with the number of requirements gives an idea of which are the most demanding ones. Each coloured line in the graph joins the stakeholder coverage values obtained from the reference Pareto front $\mathbf{F_{Def_0}}$  and each of the Pareto fronts $\mathbf{F_{Def_h}}$ $h \in \{1,2, \cdots, 5\}$  associated with the different sets of definitive stakeholders selected. It can be observed that the coverage values of $\mathbf{F_{Def_1}}$ (i.e. red line) can be clearly distinguished; they are located in the inner zone and are outnumbered by the rest. For the coverage of the rest of the fronts, we cannot see a clear separation between the lines that represent them. Both situations support the results of the Wilcoxon signed-rank test, which indicate that there is no clear difference between the coverage calculated using all demanding stakeholders in RALIC or by selecting sets of definitive stakeholders using clustering.

\begin{figure}
   \centering
    \includegraphics[width=0.99\textwidth]{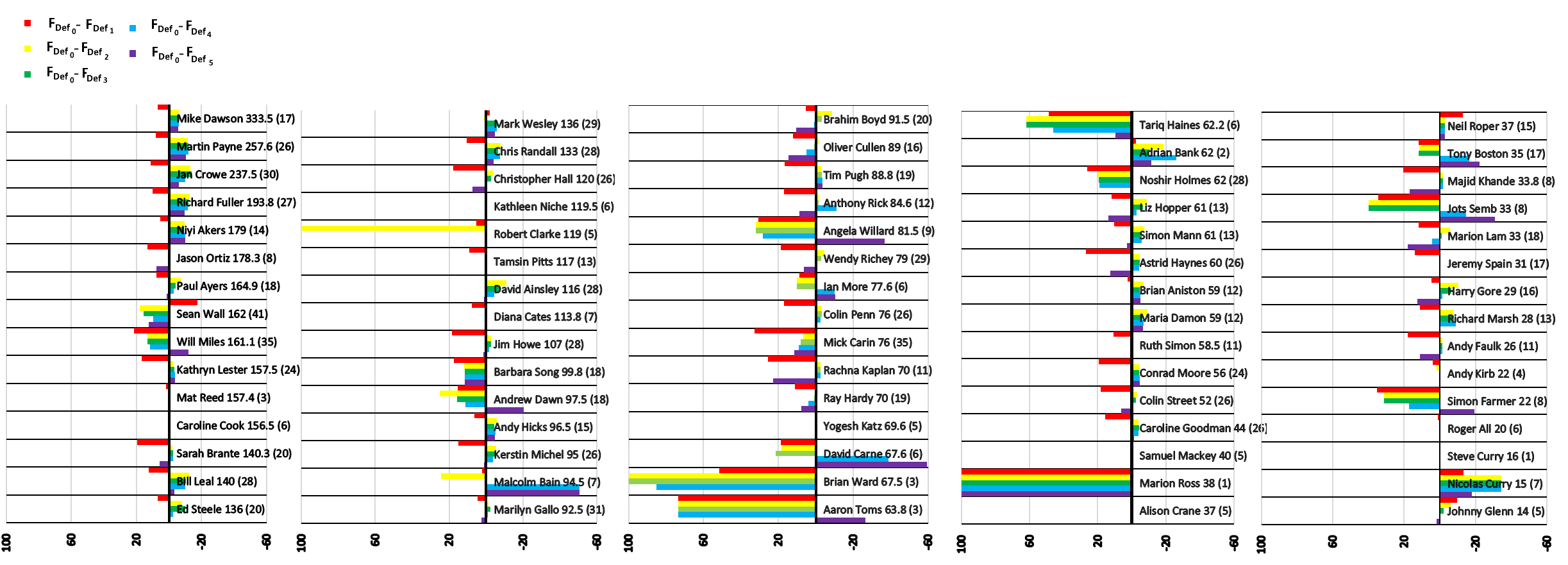}
   \caption{Differences between the coverage of demanding stakeholders, $\text{cov}_i (\mathbf{F_{Def_0}}), i \in Dem$, in RALIC and the coverage $\text{cov}_i (\mathbf{F_{Def_h}}), h \in \{1, 2, \cdots, 5\},$ obtained with the different definitive stakeholder selection methods.}

   \label{fig:differencies}
\end{figure}

At this point, it is worth noting some specific situations, such as that of "Barbara Song" and "Tariq Halnes". They get better coverage when all stakeholders are managed in the requirement selection. But they are less salient than those that are included in the first quadrant, which achieve better coverage in the Pareto fronts obtained when the set of definitive stakeholders is identified by clustering. To allow a better understanding of these and other situations, figure \ref{fig:differencies} shows the difference between the stakeholder coverage values obtained when considering all demanding stakeholders $\mathbf{Def_0}$ are considered and those obtained using the different sets of definitive stakeholders $\mathbf{Def_h}, h \in \{1,2, \cdots, 5\}$. In this graph, stakeholders are shown ordered by salience. The salience value is shown next to the stakeholder's name, and the number of requirements she demands in parentheses. The vertical line indicates that the coverage values coincide, while the bars on the left indicate that the coverage obtained by using only the selected set of definitive stakeholders is lower than the reference coverage. If the bars are to the right of the line, the coverage is higher than the baseline. Looking more closely at figure \ref{fig:differencies}, it can be seen that for most of the important stakeholders (from "Mike Dawson to "Niyi Ayers") the coverage values obtained using our proposal are slightly better. However, on some individuals  (such as  "Aaron Toms", "Brian Wars" or "Marion Ros") the coverage values are low when compared to those of other stakeholders with similar salience (such as "Colin Pen" or "Conrad Moore"). One possible explanation for this fact is that they are all outside the group of definitive stakeholders, and those with the worst coverage values demand few requirements.

In summary, we can provide a positive answer to RQ3, as there are no significant differences between the coverage values obtained when using all demanding stakeholders, or only the group of definitive stakeholders identified by using clustering techniques. Thus, our goal is fully achieved, if requirements selection is made using only the group of definitive stakeholders 
a similar overall stakeholders' coverage can be reached, 
but managing less of data.

\section{Limitations and threats to validity}

The strengths of our proposal are discussed next, considering the threats we faced along the empirical study we have conducted~\cite{wohlin2012}, also describing the measures to alleviate these threats.

\textbf{External Validity}. This kind of threat is the one that most affects our proposal, i.e. the degree to which our findings can be extrapolated to other collective of stakeholders, as we use only one data set. The difficulty lies in the need for the data set to incorporate both stakeholder data and requirement data. Stakeholder data should allow the extraction of salience components, and requirement data should incorporate individual stakeholder assessments. To our knowledge, RALIC is the only available project that meets these conditions. This could potentially lead to an external threat to our findings. However, to mitigate it, a methodological task flow has been proposed, which, when new data sets meeting the above conditions become available, can be used while ensuring replicability.

\textbf{Construct Validity}. 
These threats  indicate whether the observations used adequately capture the concepts they need to represent. The first refers to our estimates about power, legitimacy, and urgency, which have been derived from the  
available data records. Being this adaptation of the hand a possible threat, it has been based, respectively,  on the ground truth for power, on a recommendation network for legitimacy (which has been validated in several works~\cite{ZHANG2013, lim2013, VEERAPEN20151, zanaty2021}), and simple votes count for urgency.  This mapping is adequate within the scope of our work and captures the meaning that stakeholder theory defines for each salience component ~\cite{mitchell2019stakeholder}.

Another potential threat to consider is the distribution of stakeholder demands in RALIC. The demands of the stakeholders are diverse, and the points received for a demand vary between 1 and 420, leading to an imbalance in the data set. Such situations are common in real projects, and the requirements selected for inclusion in the next version could leave some stakeholders out of coverage. The coverage measure used, which is defined as a relative value (see equation \ref{eq:cov}), alleviates this problem by downplaying the imbalance.

\textbf{Internal Validity} is related to the experimental methodology that has been applied and the degree of confidence in the causal relationships established. We underpinned confidence by setting as a reference the stakeholder coverage in the original requirement selection problem. The use of several state-of-the-art clustering methods on the values of the components that define the salience mitigates potential biases in the identification of the group of definitive stakeholders. Another threat is how to choose the optimal number of groups and to deal with it, we resort to either the number of categories suggested by Mitchell et al. ~\cite{mitchell1997} or an estimate obtained by applying the \emph{majority rule} on a set of indices computed on the partitions of a data set \cite{Charrad2014}. Once the stakeholder groups have been obtained, the one with the highest salience is selected as the definitive stakeholder group, and from it a requirements selection problem is set up and solved. The comparability of stakeholder coverage results is ensured by using the same optimisation algorithm to find solutions to this and the original requirements selection problems and by the coverage measure used. All these aspects establish a general confidence framework within which we have derived the causal relationships in this study.

\textbf{Conclusion validity} affects the degree of ability to deduce the correct conclusion. Here, a threat might arise from the optimisation algorithm applied to solve \ensuremath{\mathsf{NRP}}, as solutions in the Pareto front define stakeholder coverage, and to mitigate it, we have used the same method (i.e., a greedy approach) in all cases. Another threat is related to whether a proper analysis of stakeholder coverage has been conducted. To reduce this threat, we check the stability of stakeholder coverage results based on the coefficient of variation and use the Wilcoxon test to verify that no significant differences in coverage measures are detected.

\section{Conclusions} 

Stakeholder requests are the basis for deciding on the goal of the next software to be released. Since not all stakeholders have the same importance in a project, their quantification and prioritisation play a basic role in selecting the appropriated requirements, because their judgment is one of the main criteria in the requirements selection problem. The quantitative assessment of the importance of stakeholders has been mistakenly underestimated and has not been studied before.

Typically, the quantification of stakeholders has been done by assigning weights. However, it is not clear where these values come from, nor is there a unified framework for deriving them. To alleviate this circumstance, our proposal is to quantify stakeholder importance based on the concept of salience. This concept arises in stakeholder theory with the purpose of identifying who really counts for a given project. On this basis, the stakeholder salience value was extracted from the collected stakeholder and requirement records. Then  we have proposed a process to identify the definitive stakeholder group (the most important one) and to check if this group is sufficient to uniformly cover all stakeholders involved in a software project.  In this way,  the next release problem is downsized while maintaining stakeholder coverage.

In conducting our research, we have defined a methodology represented as a task flow that, when new data sets become available, can be used while ensuring reproducibility. This methodology drives the study we have developed on the RALIC data set, which is currently the only one that incorporates both stakeholder and requirements data.

Quantification of salience has been defined by the aggregation of power, legitimacy, and urgency, with the purpose of helping to
reduce the number of stakeholders to manage in \ensuremath{\mathsf{NRP}}. 
We begin by studying the distribution of stakeholder salience values, but arranging stakeholders using salience distribution generates coverage values that are significantly different from those obtained when considering all 
stakeholders. Therefore, we use clustering methods to identify, from the different groups, the important stakeholder group (i.e., the one with the highest salience value). Each technique used found different groups, and the number of stakeholder groups suggested experimentally was 4 or 3, which coincide with those indicated by the literature (that is, definitive, dominant, dependent, and dangerous) or (that is, critical, major, and minor), respectively. 

For each of the definitive stakeholder groups we identified, we have defined the corresponding \ensuremath{\mathsf{NRP}}, found its Pareto front, and compared the stakeholder coverage values with those obtained when considering the RALIC demanding stakeholder group. We found that there are no significant differences in the stakeholder coverage values obtained. Therefore, the use of only the group of important stakeholders has proved that similar stakeholder coverage results can be reached, with a reduction between 66.32 \% and 87.75 \% in the number of stakeholders considered.

To sum up the findings, the clustering methods are useful to identify the group of important stakeholders in a project  using a quantitative definition of salience.  The number of different stakeholder groups identified matches that of the stakeholder theory and the stakeholder coverage values in the requirement selection problem hold, at least for the important ones. However, these conclusions should be taken with caution, as RALIC is the only project available to conduct this kind of study.  
For future work, we are considering some proposals. First, we plan to experiment with more sophisticated optimisation algorithms to solve \ensuremath{\mathsf{NRP}}s. In addition,  alternative data sets could be created from open source data projects, where issues can be considered as requirements, and collaborators would be stakeholders; however, figuring out the metrics is the most difficult task in this line of work. Last but not least, we wanted to study and quantify whether the proposal is beneficial or detrimental to the whole software process by collecting and studying appropriate metrics, the lack of benchmarks being a major challenge in this topic.

\end{document}